\documentclass[aps,prl,twocolumn,superscriptaddress,footinbib,floatfix,showpacs]{revtex4}

\usepackage{graphicx}

\begin{document}

\title{The Quantum Emergence of Chaos
%       \vbox to 0pt{\vss
%                    \hbox to 0pt{\hskip-50pt\rm LA-UR-00-XXXX\hss}
%                    \vskip 25pt}
}
\preprint{LA-UR-00-XXXX}
\author{Salman Habib}
\affiliation{MS B285, Theoretical Division, The University of California, 
Los Alamos National Laboratory, Los Alamos, New Mexico
87545} 

\author{Kurt Jacobs}
\affiliation{MS B285, Theoretical Division, The University of California, 
Los Alamos National Laboratory, Los Alamos, New Mexico
87545} 
\affiliation{Centre for Quantum Computer Technology, Centre for
Quantum Dynamics, School of Science, Griffith University, Nathan 4111,
Australia} 

\author{Kosuke Shizume}
\affiliation{Institute of Library and Information Science, University
of Tsukuba, 1-2 Kasuga, Tsukuba, Ibaraki 305-8550, Japan}

\begin{abstract}

The dynamical status of isolated quantum systems, partly due to the
linearity of the Schr\"odinger equation is unclear: Conventional
measures fail to detect chaos in such systems. However, when quantum
systems are subjected to observation -- as all experimental systems
must be -- their dynamics is no longer linear and, in the appropriate
limit(s), the evolution of expectation values, conditioned on the
observations, closely approaches the behavior of classical
trajectories. Here we show, by analyzing a specific example, that
microscopic continuously observed quantum systems, {\em even far from
any classical limit}, can have a positive Lyapunov exponent, and thus
be truly chaotic.

\end{abstract}


\maketitle

There can be no chaos in the dynamics of isolated or closed quantum
systems, a result which follows primarily from the linearity of the
Schr\"odinger equation~\cite{b1} and the linear Hilbert space
structure of the theory which, by virtue of the uncertainty principle,
prevents the formation of fine-scale structure in phase space
precluding chaos in the sense of classical trajectories. This leads to
a widely recognized difficulty, as classical mechanics, which
manifestly exhibits chaos, must emerge from quantum mechanics in an
appropriate macroscopic limit~\cite{b3}. The key to the resolution of
this apparent paradox lies in the fact that all experimentally
accessible situations necessarily involve measured, open systems: the
central importance of such situations in the context of chaos was
first emphasized by Chirikov~\cite{b4}. In a closely connected
development, continuous quantum measurement theory~\cite{b5} has led
to the successful understanding of the emergence of classical dynamics
from the underlying quantum physics~\cite{b8,b12,b13,b14}, and
inequalities have been derived that encapsulate the regime under which
classical motion, and thus classical chaos, exists~\cite{b12}. The
transition to classical mechanics results from the localization of the
quantum density matrix due to the information continuously provided by
the measurement (itself mediated by an environmental interaction), and
the balancing of this against the unavoidable noise from the quantum
backaction of the measurement.  For a macroscopic system, the
Ehrenfest theorem holds as a result of localization and,
simultaneously, the backaction noise is negligible, resulting in a
smooth classical trajectory.

While it has been established that observed quantum systems can be
chaotic when they are macroscopic enough that classical dynamics has
emerged, can they be chaotic outside this limit? This is the question
we address here. By defining and computing the Lyapunov exponent for
an observed quantum system deep in the quantum regime, we are able
show that the system dynamics is chaotic. Further, the Lyapunov
exponent is not the same as that of the classical dynamics that
emerges in the classical limit. Since the quantum system in the
absence of measurement is not chaotic, this chaos must emerge as the
strength of the measurement is increased, and we examine the nature
of this emergence.

The rigorous quantifier of chaos in a dynamical system is the maximal
Lyapunov exponent~\cite{b21}. The exponent yields the (asymptotic)
rate of exponential divergence of two trajectories which start from
neighboring points in phase space, in the limit in which they evolve
to infinity, and the neighboring points are infinitesimally close. The
maximal Lyapunov exponent characterizes the sensitivity of the system
evolution to changes in the initial condition: if the exponent is
positive, then the system is exponentially sensitive to initial
conditions, and is said to be chaotic. We apply this notion below to
the observation-conditioned evolution of quantum expectation values.

The evolution of a simple single-particle quantum system under an
ideal continuous position measurement is given by the nonlinear
stochastic master equation (SME) for the system density
matrix~\cite{b15}:
 \begin{eqnarray} 
  d\rho &=& - \frac{i}{\hbar} [H,\rho]dt - k[x, [x, \rho]]dt
\nonumber\\ 
        & & + 4k ( x\rho + \rho x - 2 \langle x \rangle )
                 (dy - \langle x \rangle dt)  \,,
\label{sme}
\end{eqnarray}
where the first term on the right hand side is due to unitary
evolution, $H$ being the Hamiltonian, and the second term represents
diffusion from ``quantum noise'' due to the unavoidable quantum
backaction of the measurement. The position operator is $x$, and the
parameter $k$ characterizes the rate at which the measurement extracts
information about the observable, and which we will refer to as the
{\em strength} of the measurement~\cite{b20}. The final term
represents the change in the density matrix as a result of the
information gained from the measurement. Here, $dy$ is the
infinitesimal change in the continuous output of the measuring device
in the time $dt$. The continuous output of the measuring device,
$y(t)$, referred to usually as the {\em measurement record}, is
determined by $dy = \langle x \rangle dt + dW/{\sqrt{8k}} $ where $dW$
is the Wiener increment, describing driving by Gaussian white
noise~\cite{footnote1}. The noise $dW$ is due to the fact that the
results of the measurement are necessarily random. (Note that the
backaction and $dW$ are uncorrelated with each other.) Thus on a given
experimental run, the system will be driven by a given realization of
the noise process $dW$. We will label the possible noise realizations
by $s$.

\begin{figure}
   \includegraphics[width=8.5cm,height=5.2cm]{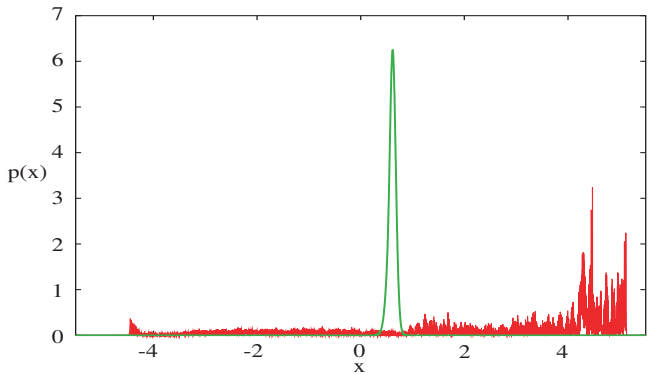}
   \caption[example]{ \label{fig0} Position distribution for
     the Duffing oscillator with measurement strengths $k=0.01$ (red)
     and $k=10$ (green), demonstrating measurement-induced
     localization ($k=10$). The momentum distribution behaves similarly.}  
\end{figure} 

A single quantum mechanical particle is in principle an infinite
dimensional system. However, for the purpose of defining an
observationally relevant Lyapunov exponent, it is sufficient to use a
single projected data stream: Here we choose the expectation value of
the position, $\langle x(t)\rangle$. The important quantity is thus
the divergence, $\Delta(t) = |\langle x(t)\rangle - \langle
x_{\mbox{\scriptsize fid}}(t) \rangle |$, between a fiducial
trajectory and a second trajectory infinitesimally close to it. It is
important to keep in mind that the system is driven by noise. Since we
wish to examine the sensitivity of the system to changes in the
initial conditions, and not to changes in the noise, we must hold the
noise realization fixed when calculating the divergence. The Lyapunov
exponent is thus
\begin{equation}
  \lambda \equiv \lim_{t\rightarrow\infty} 
  \lim_{\Delta_s(0)\rightarrow 0} \frac{\ln \Delta_s (t)}{t}
  \equiv \lim_{t\rightarrow\infty}\lambda_s(t)
\label{Lyap} 
\end{equation} 
where the subscript $s$ denotes the noise realization. This definition
is the obvious generalization of the conventional ODE definition to
dynamical averages, where the noise is treated as a drive on the
system. Indeed, under the conditions when (noisy) classical motion
emerges, and thus when localization holds (Fig.~\ref{fig0}), it
reduces to the conventional definition, and yields the correct
classical Lyapunov exponent. To combat slow convergence, we measure
the Lyapunov exponent by averaging over an ensemble of finite-time
exponents $\lambda_s(t)$ instead of taking the asymptotic long-time
limit for a single trajectory.   

A key result now follows: In unobserved, i.e., isolated quantum
dynamical systems, it is possible to prove, by employing unitarity and
the Schwarz inequality, that $\lambda$ vanishes; the finite-time
exponent, $\lambda(t)$, decays away as $1/t$~\cite{tocome}. This
theorem codifies the expectation that, since the evolution is linear,
any measure of chaos applied to it should yield a null result. As we
have emphasized earlier, however, once measurement is included the
evolution becomes nonlinear and the Lyapunov exponent need not
vanish. We now address this crucial question for a specific example.

%(Note
%that our definition of the Lyapunov exponent is drastically  different from
%other definitions of Lyapunov-like exponents for quantum systems which 
%saturate at long times, and thus do not quantify chaos in the classical
%sense of the word.) 

The system we consider is the quantum Duffing
oscillator~\cite{b26}, which is a single particle in a double-well
potential, with sinusoidal driving. The Hamiltonian for the Duffing
oscillator is    
\begin{equation}
H=p^2/2m + B x^4 - A x^2 + \Lambda x\cos(\omega t)
\label{lbham}
\end{equation}
where $p$ is the momentum operator, $m$ the particle mass, and $A$,
$B$ and $\Lambda$ determine the potential and the strength of the
driving force. We fix the values of the parameters to be $m=1$,
$B=0.5$, $A=10$, $\Lambda=10$ and $\omega=6.07$. The action of a
system relative to $\hbar$ can be varied either by changing parameters
in the Hamiltonian, or by introducing scaled variables so that the
Hamiltonian remains fixed, but the effective value of $\hbar$ becomes
a tunable parameter. Here we employ the latter choice as it captures
the notion of system size with a single number; the smaller $\hbar$
the larger the system size, and vice versa.

To examine the emergence of chaos we will first choose
$\hbar=10^{-2}$, which is small enough so that the system makes a
transition to classical dynamics when the measurement is sufficiently
strong. In this way, as we increase the measurement strength, we can
examine the transformation from essentially isolated quantum evolution
all the way to the (known) chaos of the classical Duffing
oscillator. To examine the emergence of chaos, we simulate the
evolution of the system for $k= 5\times 10^{-4}, 10^{-3}, 0.01, 0.1,
1, 10$. When $k\leq 0.01$, the distribution is spread over the entire
accessible region, and Ehrenfest's theorem is not
satisfied. Conversely, for $k=10$, the distribution is well-localized
(Fig.~\ref{fig0}), and Ehrenfest's theorem holds throughout the
evolution. Since the backaction noise, characterized by the momentum
diffusion coefficient, $D_p=\hbar^2k$, remains small, at this value of
$k$ the motion is that of the classical system, to a very good
approximation.

\begin{figure}
   \includegraphics[width=8.5cm,height=7.2cm]{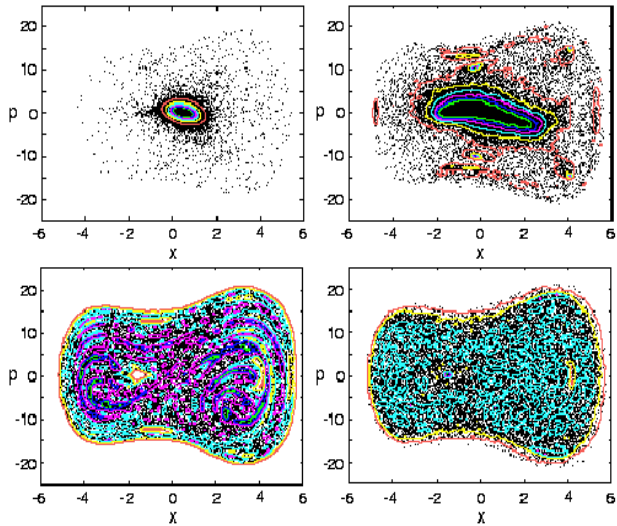}
   \caption[example]{ \label{fig1} Phase space stroboscopic maps shown
for 4 different measurement strengths, $k=5\times 10^{-4}$, 0.01
(top), and 1, 10 (bottom). Contour lines are superimposed to provide a
measure of local point density at relative density levels of
$0.05,~0.15,~0.25,~0.35,~0.45,$ and $0.55$.}  
\end{figure}

Stroboscopic maps help reveal the global structural transformation in
phase space in going from quantum to classical dynamics
(Fig.~\ref{fig1}). The maps consist of points through which the system
passes at time intervals separated by the period of the driving
force. For very small $k$, $\langle x\rangle$ and $\langle p\rangle$
are largely confined to a region in the center of phase
space. Somewhat remarkably, at $k=0.01$, although the system is
largely delocalized, as shown in Fig.~\ref{fig0}, nontrivial structure
appears, with considerable time being spent in certain outer
regions. By $k=1$ the localized regions have formed into narrower and
sharper swirling coherent structures. At $k=10$ the swirls disappear,
and we retrieve the uniform chaotic sea of the classical map (the
small `holes' are periodic islands). The swirls in fact correspond to
the unstable manifolds of the classical motion. Classically, these
manifolds are only visible at short times, as continual and repeated
folding eventually washes out any structure in the midst of a uniform
tangle.  In the quantum regime, however, the weakness of the
measurement, with its inability to crystallize the fine structure, has
allowed them to survive: we emphasize that the maps result from
long-time integration, and are therefore essentially time-invariant.

To calculate the Lyapunov exponent we implement a numerical version of
the classical linearization technique~\cite{b22}, suitably generalized
to quantum trajectories. The method was tested on a classical noisy
system with comparison against results obtained from solving the exact
equations for the Lyapunov exponents~\cite{b23}. The calculation is
very numerically intensive, as it involves integrating the stochastic
Schr\"{o}dinger equation equivalent to the SME (\ref{sme}) over
thousands of driving periods, and averaging over many noise
realizations; parallel supercomputers were invaluable for this task.

\begin{figure}
   \includegraphics[width=8.5cm,height=7.7cm]{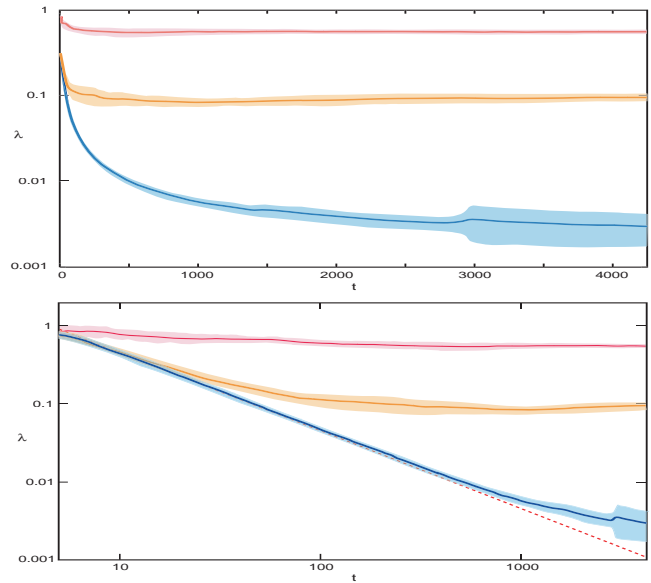}
   \caption[example]{ \label{fig2} Finite-time Lyapunov exponents
$\lambda(t)$ for measurement strengths $k=5\times 10^{-4},~0.01,~10$,
averaged over 32 trajectories for each value of $k$ (linear
scale in time, top, and logarithmic scale, bottom; bands indicate
the standard deviation over the 32 trajectories). The (analytic) $1/t$
fall-off at small $k$ values, prior to the asymptotic regime, is
evident in the bottom panel. The unit of time is the driving period.}   
\end{figure}

\begin{figure}[t]
   \includegraphics[width=8.5cm,height=5.5cm]{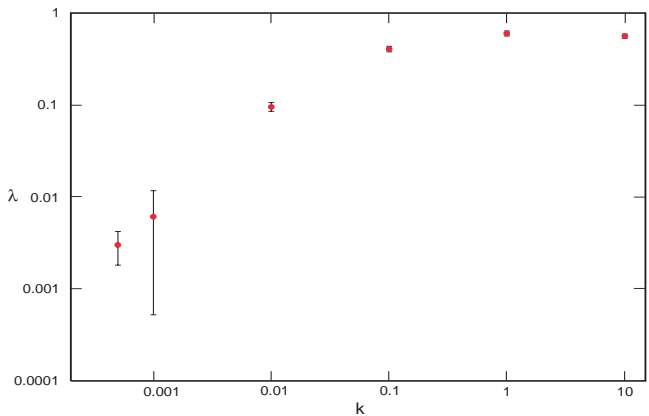}
   \caption[example]{ \label{fig3} The emergence of chaos: The
Lyapunov exponent $\lambda$ as a function of measurement strength
$k$. Error-bars follow those of Fig.~3, taken at the final time.} 
\end{figure}

We find that as $t$ is increased, for nonzero $k$, the value obtained
for $\lambda(t)$ falls as $1/t$, following the behavior expected for
$k=0$, until a point at which an asymptotic regime takes over,
stabilizing at a finite value of the Lyapunov exponent as
$t\rightarrow\infty$. This behavior is shown in Fig.~\ref{fig2} for
three different values of $k$. The Lyapunov exponent as a function of
$k$ is shown in Fig.~\ref{fig3}. The exponent increases over two
orders of magnitude in an approximately power-law fashion as $k$ is
varied from $5\times 10^{-4}$ to $10$, before settling to the
classical value, $\lambda_{Cl}=0.57$. The results in Figs.~\ref{fig2}
and \ref{fig3} show clearly that chaos emerges in the observed quantum
dynamics well before the limit of classical motion is obtained.

We now compute the Lyapunov exponent for the quantum system when its
action is sufficiently small that smooth classical dynamics cannot
emerge, even for strong measurement. Taking a value of $\hbar=16$, we
find that for $k=5\times 10^{-3}$, $\lambda=0.029\pm 0.008$, for
$k=0.01$, $\lambda=0.046\pm 0.01$ and for $k=0.02$, $\lambda=0.077\pm
0.01$. Thus the system is once again chaotic, and becomes more
strongly chaotic the more strongly it is observed. From these results,
it is clear that there exists a purely {\em quantum} regime in which
an observed system, while behaving in a fashion quite distinct from
its classical limit, nevertheless evolves chaotically with a finite
Lyapunov exponent, also distinct from the classical value. 

It is worth pointing out that an analogous analysis can also be
carried out for a continuously observed classical system. First one
notes that an {\em unobserved} probabilistic classical system also has
provably zero Lyapunov exponent: the average of $x$ for an ensemble of
classical particles does not exhibit chaos, due to the linearity of
the Liouville equation~\cite{tocome}. If we consider a noiseless
observed chaotic classical system -- possible since classical
measurements are by definition passive (no backaction noise) -- then
even the weakest meaningful measurement will, over time, localize the
probability density, generating an effective trajectory limit, and
thus the classical Lyapunov exponent,
$\lambda_{Cl}$~\cite{tocome}. Noise can always be be injected into
classical systems as an external drive, nevertheless, in the limit of
weak noise, the system will once again possess the noiseless exponent
$\lambda_{Cl}$: In a classical system the external noise is not
connected to the strength of the measurement, so one can
simultaneously have strong measurement and weak noise, which is
possible in the quantum theory only under specific
conditions~\cite{b12}. As one way to understand this case, we can
employ the quantum result as an intermediate step. Consider the
quantum Lyapunov exponent at a fixed value of $k$ (where $\lambda <
\lambda_{Cl}$) as in Fig.~\ref{fig3}. If the value of $\hbar$ is now
reduced, the dynamics of the system must tend to the classical limit
as the quantum-classical correspondence inequalities of
Ref.~\cite{b12} are better satisfied. Thus the Lyapunov exponent in
the classical limit of quantum theory -- which, to a very good
approximation, is just classical dynamics driven by weak noise -- must
tend to $\lambda_{Cl}$. If, however, the noise is not weak, an
observed classical system, like a quantum system outside the classical
regime, will also not be localized, and may well have an exponent
different from $\lambda_{Cl}$. In addition, one may expect the
non-localized quantum and classical evolutions to have quite different
Lyapunov exponents, especially when $\hbar$ is large on the scale of
the phase space, as quantum and classical evolutions generated by a
given nonlinear Hamiltonian are essentially different~\cite{Hnonlin}. 
The nature of the Lyaunov exponent for non-localized classical systems, 
and its relationship to the exponent for quantum systems is a very 
interesting open question. 

Finally, we emphasize that the chaos identified here is not merely a
formal result - even deep in the quantum regime, the Lyapunov exponent
can be obtained from measurements on a real system as in
next-generation cavity QED and nanomechanics
experiments~\cite{exp}. Experimentally, one would use the known
measurement record to integrate the SME (\ref{sme}); this provides the
time evolution of the mean value of the position. From this fiducial
trajectory, given the knowledge of the system Hamiltonian, the
Lyapunov exponent can be obtained by following the procedure described
here.

We thank Tanmoy Bhattacharya, Daniel Steck, and James Theiler for
helpful suggestions. Supercomputing resources were made available by
the LANL Institutional Computing Initiative and the Queensland
Parallel Supercomputing Facility. This work was supported by the DOE,
the ARC, and the state of Queensland.


\begin{thebibliography}{99}
\bibitem{footnote1} 
This idealized measurement has an output record with infinite
bandwidth. It is a good approximation to real measurements so long as
the bandwidth of the measuring device is large compared to that of the
system dynamics.
\end{thebibliography}
\end{document}